\documentstyle[prd,aps,epsf]{revtex}

\newcommand{\Abar}{\not{\!{\!A}}}
\newcommand{\Sbar}{\not{\!{\!S}}}
\newcommand{\pabar}{\not{\!{\!\partial}}}
\newcommand{\Od}{{\cal O}}
\newcommand{\qint}{\int d\tilde{q}}
\newcommand{\tr}{\mbox{tr}}
\newcommand{\Tr}{\mbox{Tr}}

\newcommand{\Dbar}{\not{\!{\!D}}}

\newcommand{\re}{\mbox{Re}}

\newcommand{\ZP}[1]{{\em Z.\ Phys.\ }{\bf #1}}

\begin{document}
\title{Nonlocal low-energy effective action for gravity with torsion}
\author{Antonio Dobado}
\address{Departamento de F\'{\i}sica Te\'orica, 
Universidad Complutense de Madrid, 
 28040 Madrid, Spain}
\author{Antonio L. Maroto}
\address{Astronomy Centre, University of Sussex,
               Falmer, 
               Brighton, BN1 9QJ,
               U.K.}

\date{\today}

\maketitle
\begin{abstract}
In this work we calculate the low-energy effective action for gravity 
with torsion, obtained
after the integration of scalar and fermionic matter fields, using
the local momentum representation based on the Riemann normal
coordinates expansion. By considering this expansion around different 
space-time points,
we also compute  the nonlocal terms together with the more usual 
divergent ones.
Finally we discuss the applicability of our results 
to the calculation of 
particle production probabilities.
\end{abstract}
\vskip 1.5cm


\section{Introduction}

The absence of a quantum theory of gravitation is probably one of the 
most important open
problems in theoretical physics. In fact all of our knowledge about 
gravitation refers only
to its classical aspects which are well described by General Relativity 
and other 
generalizations of this theory with the same low-energy limit. On 
the other hand, the other
interactions (strong and electroweak) are well described, even at 
the quantum level,
by the Standard Model, as it has been confirmed in detail in the 
last years at the Large
Electron-Positron Collider (LEP) and many other experiments. Moreover 
it seems 
reasonable to think that there may exist some physical regime where 
the gravitational
field could be treated classically whereas matter is quantized.
In fact this is the only regime that can be studied by now without 
making extra hypothesis
beyond what has already been checked experimentally.
In this very conservative approach one considers, on the one hand, the 
quantum gauge and matter
fields propagating on a curved space-time and, on the other hand,  
the dynamics of the 
classical degrees of freedom associated to the space-time. This is 
the so-called
 semi-classical
approximation. It is clear that space-time curvature or torsion
(see \cite{Hehl,shap3} for a review) can 
affect the quantum
fields in different ways. In addition, space-time is also affected by 
the presence of the
quantum matter fields, as it happens, for instance, in
 the electromagnetic field case, where fermion loops modify the
electromagnetic dynamics by means of the 
vacuum polarization and other 
effects. 

In order to study these new effects due to matter fields loops,  
it is specially appropriate the use of the gravitational effective action 
(EA). This
EA is obtained after integrating out the matter fields. In general 
the EA 
will be a nonlocal and non-analytical functional of the metric and
the connection. 
The exact expression for the EA, obtained after integrating out the 
matter
 fields $W[g_{\mu\nu}]$  
 is not known for arbitrary space-time geometries. However there 
are several techniques
 that have been proposed for its approximate calculation, namely:
perturbation theory, in which the metric tensor is divided in 
a flat metric and a
small fluctuation. 
The main difficulty of this method is that general covariance is explicitly lost 
\cite{thve,capp,deser}. The Schwinger-DeWitt proper time representation \cite{Schw,DeWitt}
allows us to obtain a covariant asymptotic expansion for the EA, but as far as it is local, 
it does not allow
to describe particle creation processes.
These difficulties are intended to be solved by means of the partial 
resummation 
of the Schwinger-DeWitt series \cite{Avramidi}. 
There are also perturbative methods that respect 
general covariance, 
such as the
so called covariant perturbation theory \cite{Vilkovisky}. 
Finally the local momentum representation \cite{BunchParker} is 
based on the
 Riemann normal
coordinate expansion. This technique has the advantage of combining 
the usual flat space-time
methods with formally covariant expressions. It has allowed to calculate 
the
divergent local parts of the one loop effective action for scalar and 
fermionic theories. The main aim of the present work is  to find a
representation of the {\it nonlocal finite} parts of the effective action in
this formalism.

Once we know the EA, we have all the information concerning the 
semi-classical gravitational evolution. The corresponding equations of 
motion will modify
the Einstein field equations taking into account quantum effects. Moreover, 
the EA could have
a non-vanishing imaginary part, which can be interpreted as the particle 
production probability  \cite{Schw,art2}.

In this work we show our results concerning the computation of the one-loop 
EA after 
integrating out scalar and fermionic fields by using the 
local momentum representation. Our computation includes not only the 
divergences but
also the nonlocal finite terms that can lead to instabilities of the 
classical
solutions by particle emission. As we will show, the use of normal coordinates
allow us to obtain a useful representation of  the nonlocal form factors.
Such representation has been succesfully applied in a recent work to 
obtain particle production in different cosmological contexts in a
remarkably easy way \cite{art2}. In addition, the results will
shed some light about the boundary conditions on the metric tensor which
permit a definition of the effective action.

The work is organized as follows. In Section 2 we do a brief review of 
the Euler-Heisenberg
effective lagrangian for Quantum Electrodynamics (QED). This model will 
be a guide for the
calculation of the corresponding
gravitational EA. In Section 3 we present the method to generate the
 derivative expansions
of the EA by means of Riemann normal coordinates. Applying such method to
 the scalar theory 
in the presence of gravitation, we obtain the divergences as well as the 
finite nonlocal
pieces of the EA up to quadratic terms in the curvature. In Section 4 we 
do the same with fermionic  matter fields, obtaining
the corresponding gravitational EA up to quadratic terms in the curvature and
in this case also in the torsion. In particular we apply our results to the
 Standard Model particle
 content. Finally 
 Section 5 contains
the main conclusions of this work an a brief discussion on their possible 
applications to the
computation of particle production probabilities. 
We have also included an Appendix 
containing
 the
dimensional regularization formulae and some normal coordinates expansions
 used in the text.

\section{The Euler-Heisenberg lagrangian}
The historical origin of the semi-classical EA can be traced back to the
Euler-Heisenberg lagrangian for QED \cite{EulerHeisenberg}. When the 
momentum $p$ of 
photons  is
much smaller than the electron mass $M$, the one-loop effects, such as 
vacuum
polarization, can be taken into account by adding local non-linear terms 
to the classical
electromagnetic lagrangian. Consider the QED EA given by:
\begin{eqnarray}
e^{iW[A]}=\int [d\psi][d\overline\psi]e^{-\frac{i}{4}\int d^4x 
F_{\mu\nu}F^{\mu\nu}}
\exp\left(i\int d^4x\overline \psi (i\Dbar-M+i\epsilon)\psi\right)=
e^{ -\frac{i}{4}\int d^4x F_{\mu\nu}F^{\mu\nu}} \det (i\Dbar-M+i\epsilon)
\end{eqnarray}
where as usual
 $\Dbar=\gamma^\mu(\partial_\mu-ieA_\mu)$. Using dimensional
regularization (see Appendix), it is possible to find the following 
expression up to quadratic
terms in the photon field:
\begin{eqnarray}
W[A]&=&-\frac{1}{4}\int d^4x F_{\mu\nu}F^{\mu\nu}-i\Tr \log((i\Dbar-M
+i\epsilon))
=-\frac{1}{4}\int d^4x F_{\mu\nu}F^{\mu\nu}
+i\sum_{k=1}^{\infty} \frac {(-e)^k}{k}\Tr[(i{{\pabar}} -M)^{-1} \Abar]^k 
\nonumber \\
&=&
\int d^4x \left[-\frac{1}{4}F^{\mu\nu}F_{\mu\nu}-\frac{e^2\Delta}{3(4\pi)^2}
F^{\mu\nu}F_{\mu\nu}
+\frac{2e^2}{(4\pi)^2}F^{\mu\nu}\left(\frac{2}{3}\frac{M^2}{\Box}\right. 
\right.
+\left.\left. \frac{1}{6}\left(1-2\frac{M^2}{\Box}\right)F(-\Box;M^2)\right) 
F_{\mu\nu}\right]
+\Od(A^4)
\label{ehea}
\end{eqnarray}
where $\Delta=N_\epsilon-\log (M^2/\mu^2)$, with 
$N_\epsilon=2/\epsilon+\log 4\pi -\gamma$
and $\gamma\simeq0.577$ is the Euler constant. We 
have performed the formal Taylor expansion of the logarithm and used the 
expression:
\begin{eqnarray}
F(-\Box;M^2)F_{\mu\nu}(x)=\int d^4y\frac{d^4p}{(2\pi)^4}e^{ip(x-y)}F(p^2;M^2)
F_{\mu\nu}(y)
\end{eqnarray}
with:
\begin{eqnarray}
F(p^2;M^2)=2+\int_0^1 dt \log\left(1-\frac{p^2}{M^2}t(1-t)\right)
\label{mand1}
\end{eqnarray} 
In a similar way, the inverse operator $1/\Box$ can be defined with the usual 
boundary 
conditions on the fields as:
\begin{eqnarray}
\frac{1}{-\Box}F_{\mu\nu}(x)=\int d^4y\frac{d^4p}{(2\pi)^4}e^{ip(x-y)}
\frac{1}{p^2+i\epsilon}F_{\mu\nu}(y)
\end{eqnarray}
The expression (\ref{ehea}) for the EA is nonlocal and has a regular massless limit. In fact, 
for 
small $p$ compared with $M$, the Mandelstam function $F(p^2;M^2)$ behaves as:
\begin{eqnarray}
F(p^2;M^2)= -\log \left(\frac{M^2}{-p^2-i\epsilon}\right)+\Od(M^2)
\label{mandaprox}
\end{eqnarray}
From (\ref{ehea}) we can see that the only contributions in the massless limit 
are
those coming, on one hand from the  $\Delta$ factor and, on the other hand, 
from the 
Mandelstam function. Both logarithmic contributions equal, up to  sign, 
so that they cancel each other and we obtain:
\begin{eqnarray}
W[A]=\int d^4x \left(-\frac{1}{4}F^{\mu\nu}F_{\mu\nu}
-\frac{e^2}{3(4\pi)^2} F^{\mu\nu}\Gamma(\Box)F_{\mu\nu}\right)+\Od(A^4)
\label{eheam0}
\end{eqnarray}
where we have used the following notation:
\begin{eqnarray}
\Gamma(\Box)=N_\epsilon-\log\left(\frac{\Box}{\mu^2}\right)
\label{gama}
\end{eqnarray}
to be understood as in the previous cases through the corresponding Fourier 
transform, with
the $i\epsilon$ factor as shown in (\ref{mandaprox}).
 
The EA (\ref{ehea}) allows us to derive in an exact fashion the photon 
two-point  
one loop Green functions. This, in turn, 
allows us to obtain for example the vacuum polarization. The EA can be 
expanded
as a power series in $p^2/M^2$, and also in $A$ to obtain the well-known 
Euler-Heisenberg local
lagrangian 
\cite{EulerHeisenberg}

\section{Integration of matter fields in a gravitational background}

Consider a scalar field in a curved space-time. The corresponding classical 
action is given
by:

\begin{eqnarray}
S[\phi]=-\frac{1}{2}\int d^4x\sqrt{g} \phi\left( \Box+m^2+\xi R \right) \phi
\label{acsc}
\end{eqnarray}
where
$\Box \phi=g^{\mu\nu}\nabla_{\mu}\partial_{\nu}\phi
=g^{-1/2}\partial_{\mu}\left(g^{\mu\nu}\sqrt{g}\partial_{\nu}
\phi\right)$.
The non-minimal term  $\xi R$ is included so that for  
$m=0$ and $\xi=1/6$, the classical lagrangian is invariant under
local conformal transformations.

The EA for the gravitational fields that arises after integrating out 
real scalar matter fields is given by the following expression in Lorentzian
 signature:
\begin{eqnarray}
e^{iW[g_{\mu\nu}]}=\int [d\phi]e^{iS[g_{\mu\nu},\phi]}=
  \int [d\phi]e^{-\frac{i}{2}\int d^4x\sqrt{g}\phi(\Box+m^2+\xi R 
-i\epsilon)\phi}=
(\det O)^{-1/2}
\label{eaes}
\end{eqnarray}
where  
$O_{xy}(m^2)=(-\Box_y-m^2-\xi R(y) +i\epsilon)\delta^0(x,y)$ with 
$\delta^0(x,y)$ being  
the covariant delta $\delta^0(x,y)=g^{-1/2}(x)\delta(x,y)$. Therefore we 
can write:
\begin{eqnarray}
W[g_{\mu\nu}]=\frac{i}{2}\log \; \det O(m^2)=\frac{i}{2}\Tr \;\log O(m^2)
\label{esceff}
\end{eqnarray}
Since in this expression we have only integrated the scalars out, the 
gravitational field
is treated classically. Accordingly, this EA is analogous to the classical
action but including the quantum effects due to the matter fields. In 
addition, (\ref{eaes}) 
is the generating functional of the Green functions containing only scalar 
loops and
gravitational external legs. 

Let us now consider a fermion field propagating in a curved space-time with 
torsion
\cite{Hehl}. The corresponding classical action will be given by:

\begin{eqnarray} 
S=\int d^4x \sqrt{g}\overline\psi(i\Dbar -M)\psi= \int d^4x \sqrt{g} 
\overline\psi\left[i\gamma^{\mu}\left(\partial_{\mu}-\frac{i}{2}
\Gamma^{a\;b}_{\mu}\Sigma_{ab}+\frac{i}{8}S_{\mu}\gamma_5\right)-M\right]\psi
\label{lm}
\end{eqnarray}
where $S_{\mu} = 
\epsilon_{\rho\nu\lambda\mu}T^{\rho\nu\lambda}$ 
is the torsion pseudotrace, $\Gamma^{a\;b}_{\mu}$ are the Levi-Civita 
spin-connection
components and $\Sigma_{ab}$ are the Lorentz group generators.
The fermionic EA is given by:
\begin{eqnarray} 
e^{iW[e,\hat \Gamma,A]}=
  \int [d\psi d\overline\psi]\exp \left(i\int d^4x \sqrt{g} \;\overline 
\psi (i{\Dbar} 
- M+i\epsilon)\psi\right)=\det (i\Dbar-M+i\epsilon)
\label{eafe}
\end{eqnarray} 
where $e$ denotes the vierbein, $\hat \Gamma$ the full connection
with torsion and $A$ collectively denotes the possible gauge fields. 
Accordingly:
\begin{eqnarray}
W[e,\hat \Gamma,A]=-i\log \; \det (i\Dbar-M+i\epsilon)=-i\Tr\log(i\Dbar
-M+i\epsilon)
\label{fermion}
\end{eqnarray}
In the massless limit, $M=0$, the classical fermionic lagrangian is also 
conformally invariant.
 
Since the above model does not posses self-interactions, the one-loop 
calculation is exact.
This does not mean that  it is possible 
to explicitly calculate
the EA for an arbitrary space-time geometry. In those cases in which 
there is
a high degree of symmetry, as in maximally symmetric manifolds, or in 
the so called
conformally trivial situations, i.e, conformally flat manifolds and 
conformally invariant
theories, it is possible to find the explicit form of the modified 
Einstein equations
coming from the EA \cite{BiDa}.

In order to consider more general geometries, we will use an approximation 
scheme similar 
to the one used for the Euler-Heisenberg lagrangian. It consists in treating 
the curvature
as a small perturbation. When we integrate massive fields out, this is 
equivalent to consider
that the Compton wavelength corresponding to the massive particle is much 
smaller than the
characteristic length scale of the gravitational field. In this framework, 
the expression for
the EA will be an expansion in metric tensor derivatives over the
particle mass. Such expansion will be generated by the normal coordinates 
expansions of
the  $O(m^2)$ and 
$(i{\Dbar}-M)$ operators.
In the massless case, or if we are interested in the high-energy regime, 
it is possible
to obtain an alternative expansion in powers of the curvatures (Riemann 
and Ricci tensors
and scalar curvature), generically denoted ${\cal R}$.

\subsection{Riemann normal coordinates}

Let $x^\alpha_0$ 
be the coordinates of $P$ in a given coordinate system and consider the 
set of
geodesics passing through $P$, that we will write as $x^\alpha(\tau)$.
We will choose the $\tau$ parameter in such a way that
$x^\alpha(0)=x_0^\alpha$. Each of these geodesics will be characterized 
by the tangent vector
at $P$, 
\begin{eqnarray}
\xi^\alpha=\left.\frac{dx^\alpha}{d\tau}\right\vert_{\tau=0}
\end{eqnarray}
and each point $A$ on each geodesic by certain value of the $\tau$ parameter.
The Riemann coordinates $y^\alpha$ of $A$ are defined as 
$y^\alpha=\xi^\alpha \tau$ \cite{Petrov}. In a  neighborhood of $P$ 
where any other point $A$ 
can be joined to $P$ by a unique geodesic (normal neighborhood), the 
correspondence between
the $x^\alpha$ and $y^\alpha$ coordinates is one to one.

When torsion is present and it is completely 
antisymmetric, 
the geodesic equation agrees with that obtained using
the Levi-Civita connection and therefore, some given Riemann coordinates 
respect to the Levi-Civita
connection will also be Riemannian respect to the connection with torsion 
\cite{Eisenhart,CogZer}. 
 By means of a linear real homogeneous
coordinate transformation, it is possible to write the metric tensor at $y_0$ 
in the Minkowski form
$\eta_{\mu\nu}$. 
The new coordinates are also Riemannian 
and they are known as Riemann normal coordinates.

Let us consider the components of some tensor field at $y$, that we 
will assume to be
analytic functions in a neighborhood of the normal coordinates origin $y_0$.
From their Taylor expansion around $y_0$ we can obtain 
an expression  
that is written as a series in curvatures and their covariant derivatives.  
In paricular for the metric tensor componentes we get \cite{Petrov}:
\begin{eqnarray}
g_{\mu\nu}(y)&=&\eta_{\mu\nu}+\frac{1}{3}R_{\mu\alpha\nu\beta}(y_0)y^\alpha 
y^\beta-
\frac{1}{6}R_{\mu\alpha\nu\beta ;\gamma}(y_0)y^\alpha y^\beta y^\gamma 
+
\left[\frac{1}{20}R_{\mu\alpha\nu\beta ;\gamma\delta}(y_0)
+\frac{2}{45}R_{\alpha\mu\beta\lambda}(y_0)
R^{\lambda}_{\;\gamma\nu\delta}(y_0)\right]y^\alpha y^\beta y^\gamma y^\delta
\nonumber \\
&+&\Od (\partial^5)
\label{normal}
\end{eqnarray}
Here, $\Od (\partial^5)$ denotes terms with 5 or more derivatives and
the indices in those tensors evaluated in 
$y_0$ are raised and lowered with the flat metric $\eta_{\mu\nu}$.
In the Appendix, we have written the expansions corresponding to the metric
 determinant and
other useful expressions.
It is important to note that  (\ref{normal})
is a superposition of two kind of expansions: on one hand, terms are organized
 by the
number of metric derivatives, but on the other hand, those terms with a certain
 number of
derivatives can be classified according to the number of curvature tensors they
 have.

It will also be useful to introduce the following relation:
$2\sigma(x,x')=y_\alpha y^\alpha$, 
where the biscalar $\sigma(x,x')$ represents half of the geodesic distance 
between the
$x$ and $x'$ points and 
$y^\alpha$ denotes the normal coordinates of the $x$ point with origin at $x'$.
On the other hand,
$\partial_\alpha^x\sigma(x,x')$ is a tangent vector at $x$
to the geodesic joining $x$ and $x'$, whose length equals the geodesic distance
between these two points and it is oriented in the
 $x'\rightarrow x$ direction. In turn, ${\partial_\alpha^x}'\sigma(x,x')$
is tangent to the same geodesic at $x'$, with the same modulus and oriented 
in the opposite
sense. 

In normal coordinates with origin at $x'$, according to the previous 
expressions,
we can write:
\begin{eqnarray}
\sigma_\alpha(x,x')=\frac{\partial}{\partial x^\alpha}\sigma(x,x')=y_\alpha
\end{eqnarray}
i.e, $y_{\alpha}$ are components of a vector tangent at  the origin.
 
The use of normal coordinates, apart from being basic to obtain the 
derivative expansions
of the EA, allows us to work in momentum space in a similar way
to the flat space-time. Let us consider some scalar function $f(x,y_0)$ 
with normal coordinates
with origin at $y_0$. We can define its covariant Fourier transform 
through \cite{BunchParker}:
\begin{eqnarray}
f(x,y_0)=\int \frac{d^4k}{(2\pi)^4} \hat f(k,y_0) e^{-ikx}
\end{eqnarray}
In a similar way we can introduce the covariant Dirac delta:
\begin{eqnarray}
\delta^0(x,y_0)=\int \frac{d^4k}{(2\pi)^4} e^{-ikx}
\end{eqnarray}
As far as one of the delta arguments is the origin of coordinates, we will 
have
$\delta^0(x,y_0)=\delta(x,y_0)g^{-1/2}(y_0)=\delta(x,y_0)$. In the general 
case, (arbitrary
arguments) there is also a covariant definition whose expression in arbitrary 
coordinates is
given by \cite{Avramidi}:
\begin{eqnarray}
\delta^0(x,y)=g^{1/4}(x)g^{1/4}(y)\Delta^{1/2}(x,y_0)\Delta^{1/2}(y,y_0)
\int\frac{d^4k}{(2\pi)^4}g^{-1/2}(y_0)
\exp(ik_{\mu}(\sigma^{\mu}(y,y_0)-\sigma^{\mu}
(x,y_0)))
\end{eqnarray}
where ${k_\mu}$, $\sigma_{\mu}(y,y_0)$ and $\sigma_{\mu}(x,y_0)$
are tangent vectors at $y_0$ and $\Delta(x,x')$ is the Van Vleck-Morette
determinant, defined as:
 $\Delta(x,x')=g^{-1/2}(x)\det(-\nabla_{\mu '}\nabla_\nu\sigma(x,x'))
g^{-1/2}(x')$.  
If we take $y_0$ as origin, we will have
 $\Delta(x,y_0)=g^{-1/2}(x)$, this expression reduces then to:
\begin{eqnarray}
\delta^0(x,y)=g^{-1/2}(x)
\int\frac{d^4k}{(2\pi)^4}\exp(-ik_{\mu}(x^{\mu}-y^{\mu}))
\label{delta}
\end{eqnarray}
All these definitions are valid only in normal neighborhoods of the 
origin in
which geodesics do not intersect.     

\subsection{Derivative expansions}
In normal coordinates there is a privileged point around which we 
perform the expansions.
In addition, the different curvature tensors are defined on the 
tangent plane corresponding
to that point.
This fact, together with the general coordinate invariance of the 
EA will
allow us to obtain a covariant derivative expansion for the effective 
lagrangian around
the origin. In the following we will discuss in detail the scalar case,
 although the
procedure is the same for fields with different spin.

Let us start with the scalar EA (\ref{esceff}).
Using the normal coordinates expansion for the metric tensor it is
easy to split the operator $O_{xy}(m^2)=(-\Box_y-m^2-\xi R(y) +i\epsilon)
\delta^0(x,y)$
 in a free part, that coincides with the flat space-time Klein-Gordon
operator
\begin{eqnarray}
A_{xy}(m^2)=(-\Box_0^y-m^2 +i\epsilon)\delta^0(x,y)
\end{eqnarray}
with $\Box_0^y=\eta^{\mu\nu}\partial_\mu^y \partial_\nu^y$, 
and the interaction part $B_{xy}$ that includes all the curvature dependence:
\begin{eqnarray}
B_{xy}&=&\left[-\frac{2}{3}R^\lambda_{\;\;\rho}(y_0)y^\rho
\partial_\lambda^y+\frac{1}{3}R^{\mu\;\;\nu}
_{\;\;\epsilon\;\;\beta}(y_0)y^\epsilon y^\beta \partial_\mu^y\partial_\nu^y
-\xi R(y_0)\right. 
-\left(\frac{1}{20}R_{\alpha\;\; ; \beta\gamma}^{\;\;\nu}(y_0)
+\frac{1}{20}R_{\alpha\;\; ; \gamma\beta}^{\;\;\nu}(y_0)
-\frac{1}{20}R^{\mu\;\;\nu}_{\;\;\alpha\;\;\beta;\mu\gamma}(y_0)\right.
\nonumber \\
&-&\frac{1}{20}R^{\mu\;\;\nu}_{\;\;\alpha\;\;\beta;\gamma\mu}(y_0)
 -\frac{8}{45}R_{\alpha\lambda}(y_0)R^{\lambda\;\;\nu}
_{\;\;\beta\;\;\gamma}(y_0)
+\frac{1}{15}R_{\alpha\;\;\beta\lambda}^{\;\;\mu}(y_0)
R^{\lambda\;\;\nu}_{\;\;\mu\;\;\gamma}(y_0)
+\frac{4}{45}R_{\alpha\;\;\beta\lambda}^{\;\;\mu}(y_0)
R^{\lambda\;\;\nu}_{\;\;\gamma\;\;\mu}(y_0)
+\frac{1}{40}R_{\alpha\beta;\gamma}^{\;\;\;\;\;\;\;\nu}(y_0)\nonumber \\
&+&\left.\frac{1}{40}R_{\alpha\beta;\;\;\gamma}^{\;\;\;\;\;\nu}(y_0)\right)
y^\alpha y^\beta y^\gamma \partial_\nu^y 
-\left(-\frac{1}{20}R^{\mu\;\;\nu}_{\;\;\rho\;\;\epsilon;\delta\kappa}(y_0)
+
\frac{1}{15}R_{\rho\;\;\epsilon\lambda}^{\;\;\mu}(y_0)
R^{\lambda\;\;\nu}_{\;\;\delta\;\;\kappa}(y_0)
\right) y^\rho y^\epsilon y^\delta y^\kappa 
\partial_\mu^y\partial_\nu^y\nonumber \\ 
&-&\left.\xi\frac{1}{2}
R_{;\alpha\beta}(y_0)y^\alpha y^\beta\right]\delta^0(x,y)
+... 
\end{eqnarray}
We have only written the two and four derivatives contributions since 
terms with an odd
number of metric derivatives are shown to be irrelevant for the final 
result.
Therefore we have:
$O_{xy}(m^2)=A_{xy}(m^2)+B_{xy}$

We will also assume that space-time is asymptotically  flat and this 
will allows us to
discard total derivatives of the curvatures in the EA. We have 
included the
covariant Dirac delta $\delta^0(x,y)$ in the definition of the 
free operator so that
we can use the covariant integration measure $d^4x g^{1/2}(x)$.
Taking all this into account we can write the EA as:
\begin{eqnarray}
W[g_{\mu\nu}]=\int d^4x {\cal L}_{eff}(x)=\frac{i}{2}\Tr\log O(m^2)
=\frac{i}{2}\Tr \log(A+B)
\end{eqnarray}
As far as the calculations will be done in normal coordinates with 
respect to the $y_0$ point,
any term in the effective lagrangian is evaluated in that point. For 
that reason the 
integration involved in  the $\Tr$ symbol 
cannot be done immediately. This, in turn, keep us from expanding the 
logarithm, 
which only makes sense inside the trace.
A way to avoid the problem consists in formally differentiating the EA 
with respect to  $m^2$ \cite{Zakharov}:
\begin{eqnarray}
\frac{d}{dm^2}W[g_{\mu\nu}]=-\frac{i}{2}\Tr \frac{1}{O(m^2)}=
-\frac{i}{2}\Tr((A+B)^{-1})
=-\frac{i}{2}\Tr(A^{-1}-A^{-1}BA^{-1}+A^{-1}BA^{-1}BA^{-1}-...)
\label{invdes}
\end{eqnarray}
In the last step the inverse of $A+B$ has been expanded. This can be 
done without taking the
functional trace. 

The lowest order term in $B$ is linear in the Riemann tensor and therefore 
it contains
two metric derivatives.  The term in (\ref{invdes}) with two $B$ factors
will be $\Od({\cal R}^2)$ and will contain at least four metric
derivatives, etc. Since we are only interested in the result up to 
$\Od({\cal R}^2)$, it
will be enough to consider the first three terms in (\ref{invdes}). 

\subsection{Divergent parts}

In order to calculate the divergent parts of the EA, 
we consider the previous
expansions. Since normal coordinates allows us to do momentum space 
integrations, 
we can  use the flat space-time regularization techniques such 
as dimensional
regularization. 
Our calculation method is based on the 
local momentum representation proposed in \cite{BunchParker}, but
adapted to the EA techniques. The results
agree with the well-known expressions in \cite{BiDa}. 

First we find the scalar propagator $A^{-1}_{xy}$, given by:
$A_{xy}A^{-1}_{yz}=\delta^0(x,z)$, 
where we have used the De Witt generalized summation convention for repeated 
indices.
Using (\ref{delta}), we can write this equation as:
\begin{eqnarray}
\int d^4y g^{1/2}(y)(-\Box_0^y-m^2+i\epsilon)\frac{\delta(x,y)}{g^{1/2}(x)}
 g^{-1/2}(y)\int \frac{d^4k}{(2\pi)^4} e^{-ik(y-z)}\tilde G(k)=
\frac{\delta(x,z)}{g^{1/2}(x)}
\end{eqnarray}
The $G(k)$ function is easily obtained and finally the propagator reads:
\begin{eqnarray}
A^{-1}_{yz}=\frac{1}{g^{1/2}(y)}\int \frac{d^4k}{(2\pi)^4}
e^{-ik(y-z)}\frac{1}{k^2-m^2+i\epsilon}
\label{propagador}
\end{eqnarray}
The first term in the EA expansion (\ref{invdes})
can be immediately evaluated and it reads:
\begin{eqnarray}
A^{-1}_{y_0 y_0}=\int d\tilde k \frac{1}{k^2-m^2}=-\frac{i}{(4\pi)^{D/2}}
\frac{\Gamma(1-D/2)\mu^\epsilon}{(m^2)^{1-D/2}}
\end{eqnarray}
with the notation $d\tilde k=d^Dk\mu^\epsilon/(2\pi)^D$. 
When $y_0$ appears as a repeated subindex it must be understood that the 
integration in $y_0$
has not been done. This final integration, that corresponds to the trace 
in (\ref{invdes}), 
will be performed below in an explicit way. In the last step we have used 
the equation (\ref{rdint}) from the Appendix. 
Performing the  $m^2$ integration we have the lowest order term in 
the effective lagrangian:
\begin{eqnarray}
{\cal L}_{div}^{(0)}(y_0)=\frac{1}{64\pi^2}m^4\left(\Delta+\frac{3}{2}\right)
\label{lef0}
\end{eqnarray}
where we have defined $\Delta=N_\epsilon+\log(\mu^2/m^2)$.
As is well-known, this first term will give rise to the cosmological
constant renormalization. The integration constant can be set to zero without
 loosing
generality since it can be absorbed in the renormalization procedure as we 
will show below.

Up to two metric derivatives, only the second term in (\ref{invdes}) 
contributes:
\begin{eqnarray}
(A^{-1}_{y_0 y}B_{yz}A^{-1}_{zy_0})^{(2)}&=&\int d^4y g^{1/2}(y)d^4z
g^{1/2}(z)\int d\tilde k
\frac{e^{iky}}{k^2-m^2}\left(-\frac{2}{3}R^\lambda_{\;\;\rho}(y_0)
z^\rho\partial_\lambda^z
\right.
\nonumber \\
&+& \left. \frac{1}{3}R^{\mu\;\;\nu}_{\;\;\epsilon\;\;\beta}(y_0)
z^\epsilon z^\beta \partial_\mu^z\partial_\nu^z
-\xi R(y_0)\right)\frac{\delta(y,z)}{g^{1/2}(y)}g^{-1/2}(z) 
\int d\tilde q\frac{e^{-iqz}}{q^2-m^2}
\label{2senc}
\end{eqnarray}
Integrating by parts and removing the coordinates $z$ 
through $z\rightarrow i\partial_q$, we can rewrite this expression as:
\begin{eqnarray}
(A^{-1}_{y_0 y}B_{yz}A^{-1}_{zy_0})^{(2)}&=&\int d^4z d\tilde p d 
\tilde q \frac{e^{-ipz}}{q^2-m^2}
\left(\frac{\frac{1}{3}R(y_0)-\xi R(y_0)}{(q-p)^2-m^2}
-\frac{2}{3}\frac{R^{\mu\nu}(y_0)q_\mu q_\nu}{((q-p)^2-m^2)^2}\right)
\nonumber \\
&=&
\frac{i}{(4\pi)^{D/2}}\frac{\Gamma(2-D/2)}{(m^2)^{2-D/2}}\left(\frac{1}{6}
-\xi \right) R(y_0)
\end{eqnarray}
In the last step we have done the $z$ integration first and then those 
corresponding to the
momenta. Finally integrating in $m^2$ 
we find the effective lagrangian contribution to order $\Od(\partial^2)$:
\begin{eqnarray}
{\cal L}_{div}^{(2)}(y_0)=-\frac{1}{32\pi^2}m^2\left(\Delta+1\right)
\left(\frac{1}{6}-\xi\right)R(y_0)
\label{lef2}
\end{eqnarray}

Now we derive the divergences with four metric derivatives. In this case,
 there are contributions 
from $A^{-1}BA^{-1}$ and $A^{-1}BA^{-1}BA^{-1}$:
\begin{eqnarray}
(A^{-1}_{y_0 y}B_{yz}A^{-1}_{zy_0})^{(4)}&=&\int d^4y g^{1/2}(y)d^4z
g^{1/2}(z)\int d\tilde k
\frac{e^{iky}}{k^2-m^2}\left[-\left(\frac{1}{20}
R_{\alpha\;\; ; \beta\gamma}^{\;\;\nu}(y_0)
\right.\right.
+\frac{1}{20}R_{\alpha\;\; ; \gamma\beta}^{\;\;\nu}(y_0)
-\frac{1}{20}R^{\mu\;\;\nu}_{\;\;\alpha\;\;\beta;\mu\gamma}(y_0)\nonumber \\
&-&\frac{1}{20}R^{\mu\;\;\nu}_{\;\;\alpha\;\;\beta;\gamma\mu}(y_0)
-\frac{8}{45}R_{\alpha\lambda}(y_0)
R^{\lambda\;\;\nu}_{\;\;\beta\;\;\gamma}(y_0)
+\frac{1}{15}R_{\alpha\;\;\beta\lambda}^{\;\;\mu}(y_0)
R^{\lambda\;\;\nu}_{\;\;\mu\;\;\gamma}(y_0)
+\frac{4}{45}R_{\alpha\;\;\beta\lambda}^{\;\;\mu}(y_0)
R^{\lambda\;\;\nu}_{\;\;\gamma\;\;\mu}(y_0)\nonumber \\
&+&\left.\frac{1}{40}
R_{\alpha\beta;\gamma}^{\;\;\;\;\;\;\;\nu}(y_0)
+\frac{1}{40}R_{\alpha\beta;\;\;\gamma}^{\;\;\;\;\;\nu}(y_0)\right)
z^\alpha z^\beta z^\gamma \partial_\nu^z 
+\left(\frac{1}{20}R^{\mu\;\;\nu}_{\;\;\rho\;\;\epsilon;\delta\kappa}
(y_0)-
\frac{1}{15}R_{\rho\;\;\epsilon\lambda}^{\;\;\mu}(y_0)R^{\lambda\;\;\nu}
_{\;\;\delta\;\;\kappa}(y_0)
\right)\nonumber \\
&\times& z^\rho z^\epsilon z^\delta z^\kappa \partial_\mu^z\partial_\nu^z 
-\left.\xi\frac{1}{2}
R_{;\alpha\beta}(y_0)z^\alpha z^\beta\right]\frac{\delta(y,z)}{g^{1/2}(y)}
g^{-1/2}(z)
\int d\tilde q\frac{e^{-iqz}}{q^2-m^2}
\label{2prod}
\end{eqnarray}
Removing the coordinates as before and integrating in positions and 
momenta we find:
\begin{eqnarray}
(A^{-1}_{y_0 y}B_{yz}A^{-1}_{zy_0})^{(4)}&=&-\frac{i}{(4\pi)^{D/2}}
\frac{\Gamma(3-D/2)}{(m^2)^{3-D/2}}\left(-\frac{1}{180}
R^{\mu\nu\lambda\rho}(y_0)
R_{\mu\nu\lambda\rho}(y_0)\right.
\nonumber \\
&-&\left.\frac{1}{270}R^{\mu\nu}(y_0)R_{\mu\nu}(y_0)
+\frac{1}{6}\left(\frac{1}{5}-\xi\right)\Box R(y_0)\right)
\label{42}
\end{eqnarray}
Notice that this result is finite. The divergences will appear when doing 
the 
 $m^2$ integration. We add to this term the lowest order contribution from 
the next one, 
namely:
\begin{eqnarray}
(A^{-1}_{y_0 y}B_{yz}A^{-1}_{zt}B_{tu}A^{-1}_{uy_0})^{(4)}&=&
\int d^4yg^{1/2}(y) d^4zg^{1/2}(z) d^4tg^{1/2}(t) d^4ug^{1/2}(u)
\int d\tilde k d\tilde q d\tilde p \frac{e^{iky}}{k^2-m^2}
\nonumber \\
&\times & 
\left(-\frac{2}{3}R^\lambda_{\;\;\rho}(y_0)z^\rho\partial_\lambda^z \right. 
+ \left.\frac{1}{3}R^{\mu\;\;\nu}
_{\;\;\epsilon\;\;\beta}(y_0)z^\epsilon z^\beta \partial_\mu^z\partial_\nu^z
-\xi R(y_0)\right)\frac{\delta(y,z)}{g^{1/2}(y)}g^{-1/2}(z)\frac{e^{-iq(z-t)}}
{q^2-m^2}
\nonumber \\
&\times & 
\left(-\frac{2}{3}R^\lambda_{\;\;\rho}(y_0)u^\rho\partial_\lambda^u \right.
+\left.\frac{1}{3}R^{\mu\;\;\nu}
_{\;\;\epsilon\;\;\beta}(y_0)u^\epsilon u^\beta \partial_\mu^u\partial_\nu^u
-\xi R(y_0)\right)\frac{\delta(t,u)}{g^{1/2}(t)}
g^{-1/2}(u)\frac{e^{-ipu}}{p^2-m^2}
\label{4prod}
\end{eqnarray}
In a similar fashion to the previous case we find the results for the 
regularized 
integrals:
\begin{eqnarray}
(A^{-1}_{y_0 y}B_{yz}A^{-1}_{zt}B_{tu}A^{-1}_{uy_0})^{(4)}&=&
-\frac{i}{(4\pi)^{D/2}}
\frac{\Gamma(3-D/2)}{(m^2)^{3-D/2}}\left(\frac{1}{2}\left(\frac{1}{6}
-\xi\right)^2 R(y_0)^2
-\frac{1}{108}R^{\mu\nu}(y_0)R_{\mu\nu}(y_0)\right)
\label{43}
\end{eqnarray}
Substracting  (\ref{42}) from (\ref{43}) and integrating in $m^2$ 
we obtain the divergent lagrangian up to  $\Od(\partial^4)$:
\begin{eqnarray}
{\cal L}_{div}^{(4)}(y_0)=\frac{\Delta}{32\pi^2}
\left(\frac{1}{180}R^{\mu\nu\lambda\rho}(y_0)
R_{\mu\nu\lambda\rho}(y_0)-\frac{1}{180}R^{\mu\nu}(y_0)R_{\mu\nu}(y_0)
\right.
- \left.\frac{1}{6}\left(\frac{1}{5}-\xi\right)\Box R(y_0)
+\frac{1}{2}\left(\frac{1}{6}-\xi\right)^2 R(y_0)^2\right)
\label{lef4}
\end{eqnarray}
Comparing with the well-known Schwinger-DeWitt expansion, we see that
 ${\cal L}_{div}^{(2)}(y_0)$ is proportional to $a_1(O,y_0)$ and 
${\cal L}_{div}^{(4)}(y_0)$ to $a_2(O,y_0)$.
As far as the above expressions are scalars they will have the same 
form in any coordinate
system, not neccessarily geodesic. We can then perform a coordinate change
 and integrate to obtain
the corresponding EA:
\begin{eqnarray}
W[g_{\mu\nu}]_{div}=\int d^4x {\cal L}_{div}(x)
\end{eqnarray}
We have included in ${\cal L}_{div}(x)$ the $g^{1/2}(x)$ factor coming
from the integration measure. It is possible, in principle, that when 
doing the coordinate
change, new non-covariant terms appear provided they vanish only for 
geodesic coordinates.
However, in absence of gravitational anomalies the EA is scalar and 
accordingly
such terms are not permitted. On the other hand, terms with an odd 
number of derivatives
yield terms with an odd number of momenta in the numerator 
which vanish in dimensional regularization.
The above are the only possible divergences, higher derivative terms 
give rise to 
momentum integrals with more momenta in the denominator that turn out
 to be finite. 

Apart from the divergences, which are purely local contributions, the EA
also contains finite nonlocal pieces that are responsible for the pair 
creation
processes. We have seen that in the
QED case (\ref{ehea}), the massless limit is well-defined due to the 
cancellation 
of the mass logarithmic dependence associated to the divergences with 
that coming from the nonlocal pieces. This fact allows us to extract 
some information about 
the nonlocal terms from the knowledge of the divergences. In next 
section we will profit 
this connection to find part of the nonlocal structure of the 
gravitational EA
by means of a point-splitting procedure in the divergences.

\subsection{Nonlocal contributions}

In the previous section we have obtained local contributions up to 
order $\Od({\cal R}^2)$.
If we continue the calculation to higher orders, we would get a 
power series
with terms of the following type:
\begin{eqnarray}
\frac{\nabla^n {\cal R}^p(y_0)}{m^{2p+n-4}}
\label{serie}
\end{eqnarray} 
where $\nabla$ denotes the covariant derivative. 
This is a typical derivative (or adiabatic) expansion which is only 
valid at low
energies.

All the terms in (\ref{serie})
can be classified by the number of curvatures they contain. Those terms 
with a fixed number $n$
of curvatures will give rise to a series with an increasing number of 
covariant
derivatives. If we could add all these terms together, we would obtain 
nonlocal contributions
that would provide the exact n-point Green functions (with curvatures 
in the external legs)
\cite{Ball,Avramidi,Vilkovisky}.
These Green functions will be valid for any value of the mass $m$, 
as in the QED case (\ref{ehea}).

In this section we propose a method that effectively carries out 
that resummation
for the quadratic operators, i.e we are interested in those terms
of the form $\nabla^{n} {\cal R}^2 $.
The basic idea is to perform a point-splitting in the quadratic parts
in curvatures in  (\ref{2prod}) and (\ref{4prod}), at the end of the
section we will argue that this procedure gives rise to the correct
resummation up to $\Od(m^2 {\cal R}^2)$ terms.

As we have just commented, all the terms in (\ref{serie})
are local and finite. The reason why we have not obtained nonlocal 
terms as in
(\ref{ehea}) is that the normal coordinates expansions are performed 
around a
single point $y_0$.  In (\ref{2prod}) and (\ref{4prod})
there are products of curvatures evaluated at the same point, i.e, 
 ${\cal R}(y_0){\cal R}(y_0)$.
Using again the normal coordinates expansion
we can rewrite these products as:
\begin{eqnarray}
{\cal R}(y_0){\cal R}(y_0)&=&{\cal R}(y_0){\cal R}(y)+({\cal R}(y_0)
\nabla{\cal R}(y_0)y+
{\cal R}(y_0)\nabla^2{\cal R}(y_0)yy+...)\nonumber \\
&+&({\cal R}(y_0){\cal R}(y_0){\cal R}(y_0)yy+
{\cal R}(y_0)\nabla{\cal R}(y_0){\cal R}(y_0)yyy+...)+...
\label{pointsplit}
\end{eqnarray}
This expression allows us to split the points, the price to pay is the
modification of the coefficients of the infinite higher order terms. 
We will not
modify the linear part
in ${\cal R}$ in (\ref{2prod}). Let us first obtain the nonlocal 
result and then
we will argue that the new terms $\nabla^n {\cal R}^2$ generated in
(\ref{pointsplit}) will exactly cancel those coming from the expansion
in (\ref{invdes}). 

By means of the above point-splitting, the ${\cal R}^2$ contributions in 
(\ref{2prod}) and (\ref{4prod}) will remain as:
\begin{eqnarray}
(A^{-1}_{y_0 y}B_{yz}A^{-1}_{zy_0})^{({\cal R}^2)}&=&\int d^4y g^{1/2}(y)
d^4zg^{1/2}(z)\int d\tilde k
\frac{e^{iky}}{k^2-m^2}
\left[\left(
\frac{8}{45}R_{\alpha\lambda}(y_0)
R^{\lambda\;\;\nu}_{\;\;\beta\;\;\gamma}(y)
-\frac{1}{15}R_{\alpha\;\;\beta\lambda}^{\;\;\mu}(y_0)
R^{\lambda\;\;\nu}_{\;\;\mu\;\;\gamma}(y)\right.\right.\nonumber \\
&-&\left.\left.\frac{4}{45}R_{\alpha\;\;\beta\lambda}^{\;\;\mu}(y_0)
R^{\lambda\;\;\nu}_{\;\;\gamma\;\;\mu}(y)\right)
z^\alpha z^\beta z^\gamma \partial_\nu^z 
-
\frac{1}{15}R_{\rho\;\;\epsilon\lambda}^{\;\;\mu}(y_0)
R^{\lambda\;\;\nu}_{\;\;\delta\;\;\kappa}(y)
 z^\rho z^\epsilon z^\delta z^\kappa \partial_\mu^z\partial_\nu^z
\right]\frac{\delta(y,z)}{g^{1/2}(y)}g^{-1/2}(z) \nonumber \\
&\times &
\int d\tilde q\frac{e^{-iqz}}{q^2-m^2}+\Od(\nabla^2{\cal R}^2)
\end{eqnarray}
and:
\begin{eqnarray}
(A^{-1}_{y_0 y}B_{yz}A^{-1}_{zt}B_{tu}A^{-1}_{uy_0})^{({\cal R}^2)}&=&
\int d^4yg^{1/2}(y) d^4zg^{1/2}(z) d^4tg^{1/2}(t) d^4ug^{1/2}(u)
\int d\tilde k d\tilde q d\tilde p\frac{e^{iky}}{k^2-m^2} 
\frac{e^{-ipu}}{p^2-m^2} \nonumber \\
&\times &
\left(-\frac{2}{3}R^\lambda_{\;\;\rho}(y_0)z^\rho\partial_\lambda^z
+\frac{1}{3}R^{\mu\;\;\nu}
_{\;\;\epsilon\;\;\beta}(y_0)z^\epsilon z^\beta \partial_\mu^z
\partial_\nu^z
-\xi R(y_0)\right)\frac{\delta(y,z)}{g^{1/2}(y)}g^{-1/2}(z)
\frac{e^{-iq(z-t)}}{q^2-m^2}\nonumber \\
&\times& 
\left(-\frac{2}{3}R^\lambda_{\;\;\rho}(t)u^\rho\partial_\lambda^u
+\frac{1}{3}R^{\mu\;\;\nu}
_{\;\;\epsilon\;\;\beta}(t)u^\epsilon u^\beta \partial_\mu^u\partial_\nu^u
-\xi R(t)\right)\frac{\delta(t,u)}{g^{1/2}(t)}g^{-1/2}(u)
+\Od(\nabla^2{\cal R}^2)\label{r21}
\end{eqnarray}
where, as mentioned before, $\Od(\nabla^2{\cal R}^2)$ denotes
local finite terms with two curvatures and an arbitrary even number of
derivatives. There are in principle different ways of performing the 
splitting, depending on
the choice of the pair of points, but all them are equivalent up to 
higher order terms
as can be seen from (\ref{pointsplit}).

Removing the explicit coordinates occurrences, we obtain:
\begin{eqnarray}
(A^{-1}_{y_0 y}B_{yz}A^{-1}_{zt}B_{tu}A^{-1}_{uy_0})^{(R^2)}&=&
\int d^4t d\tilde q d\tilde p e^{ip t}\left(\frac{\left(\frac{1}{3}
-\xi\right)R(y_0)}
{(q^2-m^2)^2}-
\frac{2}{3}R^{\mu\nu}(y_0)\frac{q_\mu q_\nu}{(q^2-m^2)^3}\right)
\nonumber \\
&\times& \left(\frac{-\xi R(t)}{((p+q)^2-m^2)}
+\frac{2}{3}\frac{R^{\mu\nu}(t)(p+q)_\mu (p+q)_\nu}{((p+q)^2-m^2)^2}
 \right)
+\Od(\nabla^2{\cal R}^2)
\end{eqnarray}
Using the equations (\ref{rd1}), (\ref{rd2}) and (\ref{rd4}) from the
 Appendix, 
neglecting higher order terms and
integrating in $m^2$, we obtain the following contributions to 
the effective lagrangian:
\begin{eqnarray}
{\cal L}_{1}^{({\cal R}^2)}(y_0)
=\frac{1}{32\pi^2}\int d^4t d\tilde p e^{ip t}
\left(\Delta-F(p^2;m^2)\right)
\left[\frac{1}{2}\left(\frac{1}{6}-\xi \right)^2R(y_0)R(t)
-\frac{1}{108}R_{\mu\nu}(y_0)R^{\mu\nu}(t)\right]+\Od(\nabla^2{\cal R}^2)
\label{l1}
\end{eqnarray}
$F(p^2;m^2)$ being given in (\ref{mand1}).

Apart from the local divergent and nonlocal finite terms, in the 
dimensional regularization
procedure, finite local terms do arise. However, their coefficients 
will be absorbed in the
definition of the renormalized parameters 
and they will not be explicitly considered.

In the same form as before we obtain from (\ref{r21}):
\begin{eqnarray}
{\cal L}_{2}^{({\cal R}^2)}(y_0)
=\frac{1}{32\pi^2}\int d^4t d\tilde p e^{ip t}
\left(\Delta-F(p^2;m^2)\right)
 \left[\frac{1}{180}R^{\mu\nu\lambda\rho}(y_0)
R_{\mu\nu\lambda\rho}(t)
+\frac{1}{270}R^{\mu\nu}(y_0)R_{\mu\nu}(t)\right]+\Od(\nabla^2{\cal R}^2)
\label{l2}
\end{eqnarray}
Adding both contributions  in (\ref{l1}) and 
(\ref{l2}) and including the $0$ and $2$ derivatives divergent contributions 
given in
 (\ref{lef0}) and (\ref{lef2}), we can write the nonlocal EA in a slightly
different notation:
\begin{eqnarray}
W[g_{\mu\nu}]&=&\frac{1}{32\pi^2}\int d^4x \sqrt{g}\left[
\frac{m^4}{2}\left(\Delta+\frac{3}{2}\right) 
-m^2\left(\Delta+1\right)
\left(\frac{1}{6}-\xi\right)R(x)
+\Delta 
\left(\frac{1}{180}R^{\mu\nu\lambda\rho}(x)
R_{\mu\nu\lambda\rho}(x)
\right.\right.\nonumber \\
&-&\left.\frac{1}{180}R^{\mu\nu}(x)R_{\mu\nu}(x)
+ \frac{1}{2}\left(\frac{1}{6}-\xi\right)^2 R(x)^2\right)
-\left(\frac{1}{180}R^{\mu\nu\lambda\rho}(x)F(\Box;m^2)
R_{\mu\nu\lambda\rho}(x)\right. 
\nonumber \\
&-&\left. \left. \frac{1}{180}R^{\mu\nu}(x)F(\Box;m^2)R_{\mu\nu}(x)
+\frac{1}{2}\left( \frac{1}{6}-\xi\right)^2 R(x)F(\Box;m^2)R(x)\right)
+ \Od(m^2{\cal R}^2)\right] +\Od ({\cal R}^3)
\label{ear2}
\end{eqnarray}
Here the $F(\Box;m^2)$ operator action should be understood through 
the expressions
 (\ref{l1}) y (\ref{l2}) as we did in the QED case (\ref{ehea}), 
in addition we have neglected local finite 
terms and total derivatives. 
On the other hand, we see that the quadratic
divergences agree with those obtained in the previous section.

Consider now the massless limit of the EA in (\ref{ear2}). 
For small masses compared with $p$, the Mandelstam function behaves as 
shown in
(\ref{mandaprox}). 
As in the QED case, the term  proportional to 
 $\log (m^2)$, associated to the divergences in (\ref{ear2}),
has the same coefficient but with opposite sign as that coming from 
the Mandelstam function.
Therefore they exactly cancel. This allows us to obtain the regular 
massless limit:
\begin{eqnarray}
W[g_{\mu\nu}]&=&\frac{1}{32\pi^2}\int d x \sqrt{g}
\left(\frac{1}{180}R^{\mu\nu\lambda\rho}(x)\Gamma(\Box)
R_{\mu\nu\lambda\rho}(x)-\frac{1}{180}R^{\mu\nu}(x)\Gamma(\Box)
R_{\mu\nu}(x)
\right. \nonumber \\
&+& \left. 
\frac{1}{2}\left(\frac{1}{6}-\xi\right)^2 R(x)\Gamma(\Box)R(x)\right)
+\Od ({\cal R}^3)
\label{nolocal}
\end{eqnarray}  
where $\Gamma(\Box)$ is given in (\ref{gama}).
This result is what one had expected on dimensional grounds \cite{Duff}, 
however we have got in addition an explicit representation, completely
analogous to the flat space-time case, for
the form factors. 
In principle, the result in (\ref{ear2}) cannot be considered, 
strictly speaking,  as the 
 $\Od({\cal R}^2)$ contribution in the curvature expansion since in 
the calculation
we have not considered those terms with two curvatures and an arbitrary 
number of covariant
derivatives, denoted by  
$\Od(\nabla^2{\cal R}^2)$. We have obtained a nonlocal expression, 
but the infinite $\Od(\nabla^2{\cal R}^2)$ terms could in principle modify it.
However, we have seen in the previous section that the divergences
have a logarithmic dependence in $m^2$ (whose calculation is unambiguous).
In addition, we showed in Section 2 in the QED case that the coefficients of 
the $\log(m^2)$ in the divergent part have to be the same as those
of the $\log(p^2)$ in the nonlocal pieces, in order to have a regular
masless limit. In (\ref{nolocal}) we have seen that our result indeed
posseses a regular masless limit. Since the $\Od(\nabla^2{\cal R}^2)$ terms
are finite, they could only modify the finite local and nonlocal part, 
but not the
divergences. However such a modification in the nonlocal part could
spoil the regularity of the masless limit. Accordingly, we conclude that
the $\Od(\nabla^2{\cal R}^2)$ terms can only modify the local finite 
pieces and those 
proportional to $m^2$, but not the nonlocal ones containing $\log (m^2)$.
For that reason, the result in (\ref{ear2}) includes all the quadratic 
curvature contributions
except for those  containing an arbitrary mass power denoted 
$\Od(m^2{\cal R}^2)$.

A compatible result has been obtained using partial resummation of 
the Schwinger-DeWitt expansion
\cite{Avramidi} and also by means of the so called covariant 
perturbation theory 
\cite{Vilkovisky}, in this case it was possible to derive the 
cubic terms in curvatures for asymptotically flat manifolds.
Notice however that the boundary conditions that normal coordinates 
impose on
the metric tensor are slightly more general  than the asymptotically
Minkowskian condition. In fact, we have only required that the curvatures
and their covariant derivative vanish at infinity. This condition 
includes, for instance, those Robertson-Walker manifolds in which
the expansion rate $\dot a/a$ asymptotically vanishes
in the future,  although the scale
factor itself $a(t)$ does not tend to a constant. This kind of manifolds
are known to accept the definition of the so-called adiabatic
vacua \cite{BiDa}. In \cite{art2} the possibility of defining
such vacua was extremely useful for the interpretation of the
effective action as vacuum persistence amplitude.  

To summarize, the procedure we have just presented makes it possible 
a partial resummation
of the higher order terms in the EA, within the mentioned 
limits. Neglecting  $\Od({\cal R}^3)$  terms implies that
 (\ref{ear2}) will be  a good approximation for  
 $\nabla \nabla {\cal R}>>{\cal R}^2$. In a similar fashion, when we 
neglect 
$\Od(m^2{\cal R}^2)$ terms we are assuming that
$\nabla \nabla {\cal R}>>m^2 {\cal R}$. However, if we are only interested 
in the two point
Green functions with external curvature legs, the massless limit in 
(\ref{nolocal}) is exact. The renormalization of the EA 
can be done following the standard procedure in the classical references 
(\cite{BiDa} and \cite{shap3}).
Here we will only mention that from the viewpoint of the Appelquist and 
Carrazone decoupling
theorem \cite{AppCarr}, the scalar field does not decouple from 
gravity since
there are new terms in the EA (\ref{ear2}) which are not present in the 
Einstein-Hilbert action
and they are not suppressed by powers of the particle 
mass  $m$.

\section{Integration of the Standard Model matter fields: the EA for torsion}

Up to now we have only worked with scalar fields. However, in order to 
include the effect of the
matter content present in the SM in the gravitational EA, we have to 
deal with the integration
of fermionic fields.  In this case we can proceed in a similar way 
using
 (\ref{fermion}) in order to obtain the EA for the gauge fields and 
gravitation
by integrating out the SM matter fields. The main novelty is that, in 
addition to the gravitational
field (the vierbein), fermions couple also to the pseudotrace of the 
torsion as it 
was discussed at the beginning of the previous section. 
The SM matter lagrangian in a curved space-time with torsion can be 
written as
\cite{anom}:
\begin{eqnarray}
{\cal L}_M=\sqrt{g} \left(\overline {\cal Q}(i\Dbar^Q-M^Q) {\cal Q}+
\overline {\cal L}(i\Dbar^L-M^L) {\cal L}\right)
\label{materia}
\end{eqnarray}
where:
\begin{eqnarray}
\Dbar^Q  = \gamma^\mu D^Q_\mu= \gamma^{\mu}(\partial_{\mu}
+\Omega^{Q}_{\mu}
+S^{Q}_{\mu}\gamma_5), \;\;
\Dbar^L =  \gamma^\mu D^L_\mu=\gamma^{\mu}(\partial_{\mu}
+\Omega^{L}_{\mu}+S^{L}_{\mu}\gamma_5)
\label{dos}
\end{eqnarray}
with $M^Q$ and $M^L$  the mass matrices of quarks and leptons. 
We have followed the notation in \cite{anom}.
In the following we will concentrate only in the gravitational 
couplings, so that we have  
neglected the gauge fields contributions in the operators. 

In order to obtain the torsion contribution to the EA up to $\Od(S^2)$, 
we first consider a flat space-time with torsion and afterwards we will 
include the effect
of curvature. From the above lagrangian we see that torsion behaves as 
the electromagnetic
field in QED, without considering the  $\gamma_5$ coupling, which does 
not affect
the calculation of terms with an even number of fields, in the massless 
limit. Using the result in
(\ref{ehea}) and changing 
$e\Abar \rightarrow -1/8 {\Sbar} \gamma_5$, we obtain in this limit:
\begin{eqnarray}
W[S]=-\left(\frac{N_\nu}{2}+N_{Df}+N_c N_q\right)
\frac{N_f}{192(4\pi)^2}\int d^4x 
 S^{\mu\nu}\Gamma(\Box) 
S_{\mu\nu}+\Od(S^4)
\label{sea}
\end{eqnarray}
where $\Gamma(\Box)$ is defined in 
(\ref{gama}), $N_f$ is the number of families, $N_\nu$ the number of 
neutrinos, 
$N_{Df}$ that of Dirac fermions and $N_q$ 
the number of quark flavors. 
$S_{\mu\nu}=\partial_\mu S_\nu-\partial_\nu S_\mu$.
Due to the absence of right neutrinos, there could be, in principle, 
parity violating
 terms. However it can be shown that those terms  are total derivatives.

In order to introduce the space-time curvature, we recall that the 
SM matter sector is
locally conformally invariant (for massless fermions), this is also 
the case of the 
counterterms \cite{duff},
 (see \cite{shap2,shap3} where the renormalization procedure
has been studied for manifolds with torsion).
When including the curvature, apart from those in  (\ref{sea}),
there could be quadratic terms in torsion in the generic form 
${\cal R}S^2$. However such terms are not conformally invariant. 
Therefore the ones obtained
above are the only possible divergences. Next we will confirm these 
results with an explicit 
calculation of the divergences in  a space-time with curvature.

The SM Dirac operators in Euclidean space are not Hermitian because of 
the electroweak gauge couplings and the absence of the right
neutrinos. However,  in Euclidean space the EA 
divergences are real
\cite{Ball}, thus it is enough to calculate: 
$2\re\; W[e,\hat \Gamma]=-\log \det ({\cal O})$,
where ${\cal O}=(\Dbar +M)^\dagger (\Dbar +M)$ and $\Dbar$
and M denotes the joint Euclidean Dirac operator and mass matrix 
for quarks and leptons. 
The heat-kernel expansion together with dimensional regularization 
\cite{BiDa,JaOs}
allows us to obtain:
\begin{eqnarray}
-\frac{1}{2}\Tr \log ({\cal O})=\frac{\mu^{\epsilon}}
{2(4\pi)^{D/2}}\tr\sum_{n=0}^{\infty}M^{D-2n}
\Gamma\left(n-\frac{D}{2}\right)\int d^4x \sqrt{g}\;\, 
a_n({\cal O},x)
\end{eqnarray}
The well-known HMDS coefficients  $a_n$ are given for the above operators 
by:
\begin{eqnarray}
a_0({\cal O},x)&=&1, \;\;
a_1({\cal O},x)= \frac{1}{6}R-X \nonumber \\
a_2({\cal O},x)&=&\frac{1}{12}[D_{\mu},D_{\nu}][D^{\mu},D^{\nu}]
+\frac{1}{6}
[D_{\mu},[D^{\mu},X]]+ \frac{1}{2}X^2-
\frac{1}{6}RX \nonumber\\
&-&\frac{1}{30}R_{;\mu}^{\;\;\mu}+\frac{1}{72}R^2+\frac{1}{180}
(R_{\mu \nu \rho
\sigma}  R^{\mu \nu \rho \sigma} -R_{\mu \nu} R^{\mu \nu})
\label{swdwdes}
\end{eqnarray}
where, when we neglect the gauge fields contributions, the operator can
be written as:
\begin{eqnarray}
{\cal O}=D_{\mu}D^{\mu}+X+M^2,\;\;\;  \mbox{with}\;\;\;
X=\gamma_5 S^{\mu}_{\; ;\mu}+2S_{\mu}  S^{\mu}-
\frac{1}{4}[\gamma^{\mu},\gamma^{\nu}][d_{\mu},d_{\nu}]\nonumber
\end{eqnarray}

\noindent and \centerline{$D_{\mu}=\partial_\mu+\Omega_\mu
-\frac{1}{2}\gamma_5[\gamma_{\mu},
\gamma^{\nu}]S_{\nu}
=d_{\mu}
-\frac{1}{2}\gamma_5[\gamma_{\mu},\gamma^{\nu}]S_{\nu}$}

Writing the result in Lorentzian signature we have (see also the previous
works \cite{shap2,shap3}):
\begin{eqnarray}
W_{div}[e,\hat \Gamma]&=&-\sum_f \sum_i N_i\frac{1}{32\pi^2}\int 
d^4x \sqrt{g}
\left[M_i^4\left(\frac{\Delta_i}{2}+\frac{3}{4}\right)-M_i^2\left(\Delta_i
+1\right)
\left(-\frac{1}{12}R
+\frac{1}{32}S^2\right)\right.\nonumber\\ 
&+&\left.\Delta_i\left(\frac{1}{384}S^{\mu\nu}S_{\mu\nu}
-\frac{7}{1440}R_{\mu\nu\rho\sigma}R^{\mu\nu\rho\sigma}
-\frac{1}{180}R_{\mu\nu}R^{\mu\nu}
+\frac{1}{288}R^2\right)\right]
\label{sdiv}
\end{eqnarray}
$\sum_f$ denotes sum over the different families  and  $\sum_i=\sum_{lept.}
+N_c\sum_{quarks}$ over the different flavors in each family, including 
quarks and leptons. 
On the other hand,
$N_i$ is the number of spinor components: 2 for neutrinos and 4 for the
 rest of fermions.
$M_i$ are the different fermion masses, $\Delta_i=N_\epsilon
+\log (\mu^2/M_i^2)$ and
$\mu$ the renormalization scale.

This formula is compatible with the flat space-time result in
 (\ref{sea}). There is no
 ${\cal R}S^2$ term, as commented before, and we have discarded 
total derivatives. 
Following similar steps as for the scalar case, i.e. including 
$\log (\Box/M^2)$ 
factors and taking the massless limit, we obtain 
the finite nonlocal contributions depending on curvature and torsion
 from the divergences. 
In the mentioned limit we have:
\begin{eqnarray}
W[e,\hat \Gamma]&=&-\frac{N_f(8N_c+6)}{32\pi^2}\int d^4x \sqrt{g}
\left(\frac{1}{384}S^{\mu\nu}\Gamma(\Box)S_{\mu\nu}
+\frac{1}{288}R\Gamma(\Box)R
\right.\nonumber\\
&-&\left.\frac{7}{1440}R_{\mu\nu\rho\sigma}\Gamma(\Box)R^{\mu\nu\rho\sigma}
-\frac{1}{180}R_{\mu\nu}\Gamma(\Box)R^{\mu\nu}
\right)
+\Od({\cal R}^3)+\Od(S^3)
\end{eqnarray}
where $\Gamma(\Box)$ is given in (\ref{gama}) and $\Od(S^3)$ denotes terms 
with 3 or more torsion fields. Finally, due to the presence of chiral 
fermions, the EA could contain
an abnormal parity sector, responsible for the gauge and gravitational 
anomalies. However, 
as far as the previous result has been derived from the real part of the EA 
(with normal parity), this sector is not taken into account.

The renormalization procedure in this case will require, not only the 
introduction of 
quadratic terms and a constant, but also a kinetic and mass terms for
 the torsion field.
Accordingly, the starting classical action should be:
\begin{eqnarray}
S_{G}=\int d x \sqrt{g}\left(\frac{R-2\Lambda}{16\pi G}+a_2
R^{\mu\nu\lambda\rho}
R_{\mu\nu\lambda\rho}+a_3R^{\mu\nu}R_{\mu\nu}
+ a_4R^2
+a_5 S_{\mu\nu}S^{\mu\nu}+a_6S^2\right)
\end{eqnarray}
The divergences can be absorbed by constants redefinition:
\begin{eqnarray}
a_j &\rightarrow & a'_j=a^r_j(\mu)
-\frac{1}{32\pi^2}C_j N_\epsilon+\mbox{finite constants},\; j=0..6
\end{eqnarray}
where $a_0=-2\Lambda/(16\pi G)$ and $a_1=1/(16\pi G)$.
The values of  
$C_j$ are shown in Table 3.1.

The scale dependence of the renormalized constants is derived from the 
renormalization
group equation \cite{shap3,shap2}, in an analogous way to the scalar case. 
As a consequence, if at a given 
scale $\mu$, 
the constants have certain values $a_i^r(\mu)$, 
their values at a different scale $\mu'$ will be given by:
\begin{eqnarray}
a_i^r(\mu')=a_i^r(\mu)+\frac{C_i}{32\pi^2}\log\left(\frac{\mu^2}{\mu'^2}
\right)
\label{escala}
\end{eqnarray}
where the  $C_i$ constants are those appearing in Table 3.1. In particular, 
the renormalized 
Newton constant $G^r(\mu)$ has a scale 
dependence and therefore
its value should be specified for a given $\mu$. 
Thus  $G$ will depend on the size of the system we are 
considering and this
could have an enormous importance in cosmology (see \cite{goldman}).
\begin{center}
\begin{tabular}{|c|c|c|c|} \hline
$C_0$&$C_1$&$C_2$&$C_3$\\  \hline
$-\sum_f \sum_i N_i  \frac{M_i^4}{2}$&
$-\sum_f \sum_i N_i  \frac{M_i^2}{12}$&$\frac{7}{180}N_f N_c
+\frac{7}{240}N_f$&
$\frac{2}{45}N_f N_c+\frac{1}{30}N_f$\\ 
\hline \hline
$C_4$&$C_5$&$C_6$& \\ \hline
$-\frac{1}{36}N_f N_c-\frac{1}{48}N_f$&
$-\frac{1}{48}N_f N_c-\frac{1}{64}N_f$&
$\sum_f \sum_i N_i \frac{M_i^2}{32}$& \\ \hline
\end{tabular}
\end{center}
\centerline{\footnotesize {\bf Table 3.1:} Renormalization constants 
for the SM case}

In the renormalized EA we can generate a torsion kinetic term by
means of a $S_\nu$ finite normalization as follows: $S_{\nu}^r=
Z_3^{-1/2}(\mu)S_{\nu}$, 
where $a^r_5(\mu)=-\frac{1}{4}Z_3^{-1}(\mu)$. Thus in the massless 
limit we have:
\begin{eqnarray}
W^r[S^r]&=&\int d^4x \sqrt{g} \left(-\frac{1}{4}S_{\mu\nu}^r S^{r\mu\nu}
+\mbox{nonlocal terms} +
\Od(S^3)\right)
\end{eqnarray}
As a consequence the physical torsion field
$S_\mu^r$ will behave as an abelian gauge field. 
Therefore we have generated a kinetic term for
torsion even starting from a theory without propagating torsion.

\section{Conclusions and discussion}

In this work we have dealt with the computation of the low-energy 
Effective Action (EA)
for gravity obtained when matter fields, both scalar and fermionic, 
are integrated out.
As a much simpler exercise, we have started with by reviewing the
 low-energy
EA for the electromagnetic field obtained when the electronic field is
integrated out. Special attention has been paid to the nonlocal terms 
which are related
to the particle production probabilities.   
 
In order to integrate the scalar fields to compute the EA for gravity,
 we have used the
normal coordinate expansion. This method has 
been used 
previously to compute the divergent terms of the EA. By comparison of 
this kind of
expansion around different space-time points, we have been able to 
extend the
previous work to obtain also the nonlocal finite terms of the EA up 
to quadratic terms 
in the curvature. In particular we have arrived to the 
conclusion that
the scalar field does not decouple from the EA in the large mass limit 
in the
 Appelquist-Carrazone sense. Finally we have also discussed  the
 meaning of 
the expansion in the massless limit. 
 
In order to be able to consider the matter field content present in
 the Standard Model
we have also studied the case of fermionic matter. The main novelty 
in this case is that
this kind of fields can also couple to the torsion pseudotrace in 
addition to the
gravitational field. Thus we have
obtained the low-energy EA for the gravitational field and the 
torsion including the
divergent and the nonlocal finite terms up to quadratic terms 
in the curvature and the
torsion. We have discussed the renormalization of  
this EA and we have found 
that a kinetic term is
generated for the torsion pseudotrace. This result is quite 
interesting since in the
standard Einstein-Cartan action \cite{Hehl}, the torsion does not propagate. 

Finally we would like to stress that the results obtained in this 
work can be useful
for the study of some interesting physical effects. In particular, 
the low-energy EA
can be applied for the study of the quantum stability of classical 
solutions of the
Einstein equations of motion. More specifically, the nonlocal terms 
of the EA
which have been computed here, are complex in general.  Classical 
solutions that give
rise to an imaginary part when the EA is evaluated on them, are 
unstable by particle
radiation even if they are stable at the classical level. Moreover,
 the imaginary part
of the EA evaluated on classical solutions, can be used to compute 
particle production
rates and spectra thus  providing an alternative method to the more
 commonly used based
on the Bogolyubov transformations. The expression that we have obtained
for the nonlocal form factors is particularly appropriate for
the calculation of particle production probabilities in cosmological
space-times. The boundary conditions imposed by the normal coordinates
expansion, allow to apply this method not only to asymptotically
flat manifolds, but also to some other cosmological manifolds where
the expansion rate asymptotically vanishes.
Work has been done in this
 direction and it  will be
presented elsewhere \cite{art2}.

\vspace{0.5cm}

{\bf Acknowledgments:}
We thank I.L.Shapiro for his comments and suggestions in several
points of the paper. 
This work has been partially supported by the Ministerio de Educaci\'on y
Ciencia (Spain)(CICYT AEN96-1634). A.L.M. acknowledges support from
SEUID-Royal Society.

\section{Appendix}
\subsection*{Dimensional regularization formulae}
In dimensional regularization \cite{regdim}, the space-time dimension is 
taken to be
$D=4-\epsilon$ and the poles are parametrized through
$N_\epsilon$ that was defined in Section II. We will use the notation
$d\tilde q=\mu^\epsilon d^Dq/(2\pi)^D$ where $\mu$ is the renormalization 
scale.
Some useful expressions are:
\begin{equation}
\qint\frac{(q^2)^r}{[q^2-R^2]^m}=
i\frac{(-1)^{r-m}}{(4\pi)^{D/2}}\frac{\Gamma(r+D/2)
 \Gamma(m-r-D/2) \mu^{\epsilon}}{\Gamma(D/2)\Gamma(m)(R^2)^{m-r-D/2}}
\label{rdint}
\end{equation}
\begin{eqnarray}
I_1(p,m^2)\equiv \int d\tilde q\frac{1}{(q^2-m^2)^2((q+p)^2-m^2)}
&=&
i\frac{(-1)^{D/2}}{(4\pi)^{D/2}}\Gamma(3-D/2)\int_0^1 
dt (1-t)(-m^2+p^2t(1-t))^{D/2-3}
\nonumber \\
\int^{m^2} I_1(p,m^2)dm^2 &=&\frac{i\mu^{\epsilon}}{2(4\pi)^2}
\left(N_\epsilon+2-\log \frac{m^2}{\mu^2}
-F(p^2;m^2)\right)
\label{rd1}
\end{eqnarray}

\begin{eqnarray}
I^{\mu\nu}_2(p,m^2)&\equiv&\int d\tilde q\frac{q^\mu q^\nu}{(q^2-m^2)^3((q+p)^2
-m^2)}\nonumber \\
&=&i\frac{(-1)^{D/2}}{2(4\pi)^{D/2}}\Gamma(4-D/2)\int_0^1 
dt (1-t)^2(-m^2+p^2t(1-t))^{D/2-4}
 \left(p^\mu p^\nu+g^{\mu\nu}\frac{-m^2+p^2
 t(1-t)}{2(3-D/2)}\right)\nonumber \\ 
\int^{m^2}I_2^{\mu\nu}(p,m^2)dm^2 &=&\frac{i\mu^\epsilon}{12(4\pi)^2}
\left(N_\epsilon+\frac{13}{6}
-\log \frac{m^2}{\mu^2}-F(p^2;m^2)\right)g^{\mu\nu}
+\Od(p^2)
\label{rd2}
\end{eqnarray}

\begin{eqnarray}
I^{\mu\nu\alpha\beta}_3(p,m^2)&\equiv&\int d\tilde q
\frac{q^\mu q^\nu q^\alpha q^\beta}
{(q^2-m^2)^4((q+p)^2-m^2)}\nonumber \\
&=&i\frac{(-1)^{D/2}}{6(4\pi)^{D/2}}\Gamma(5-D/2)\int_0^1 
dt (1-t)^3(-m^2+p^2t(1-t))^{D/2-3}
\frac{g^{\alpha\nu}g^{\beta\mu}+g^{\alpha\mu}
g^{\beta\nu}+g^{\alpha\beta}g^{\mu\nu}}
{(6-D)(8-D)}\nonumber \\
&+&\Od(p^2)\nonumber \\ 
\int^{m^2}I_3^{\mu\nu\alpha\beta}(p,m^2)dm^2 &=&
\frac{i\mu^\epsilon}{96(4\pi)^2}\left(N_\epsilon+\frac{7}{3}
-\log \frac{m^2}{\mu^2}-\left(1-\frac{2m^2}{p^2}\right)F(p^2;m^2)-
\frac{4m^2}{p^2}\right)\nonumber \\
&\times &(g^{\alpha\nu}g^{\beta\mu}+g^{\alpha\mu}g^{\beta\nu}
+g^{\alpha\beta}g^{\mu\nu})+\Od(p^2)
\label{rd4}
\end{eqnarray}
where $F(p^2;M^2)$ is given in (\ref{mand1}). 
These expressions have been obtained in the Minkowski space-time. 
We have not explicitly 
included the $+i\epsilon$ terms accompanying the momenta in order 
to avoid the
confusion with that coming from dimensional regularization.

\subsection*{Normal coordinates expansions}

We will show the Riemann normal coordinates expansions for 
different objects up to
order $\Od(\partial^4)$. We will take the point $y_0$ as the 
coordinates origin.
For the inverse metric tensor we have:
\begin{eqnarray}
g^{\mu\nu}(y)&=&\eta^{\mu\nu}-\frac{1}{3}
R^{\mu\;\;\;\nu}_{\;\;\alpha\;\;\beta}(y_0)y^\alpha y^\beta
+\frac{1}{6}R^{\mu\;\;\;\nu}_{\;\;\alpha\;\;\beta ;\gamma}(y_0)
y^\alpha y^\beta y^\gamma
+
\left[-\frac{1}{20}R^{\mu\;\;\;\nu}_{\;\;\alpha\;\;\beta ;\gamma\delta}
(y_0)
+\frac{1}{15}R_{\alpha\;\;\;\beta\lambda}^{\;\;\mu}(y_0)
R^{\lambda\;\;\nu}_{\;\gamma\;\;\delta}(y_0)\right]
y^\alpha y^\beta y^\gamma y^\delta \nonumber \\
&+&\Od (\partial^5)
\end{eqnarray}

The metric determinant $g=\vert\det g_{\mu\nu}\vert$ has the 
following expansion:
\begin{eqnarray}
g(y)=1+\frac{1}{3}R_{\alpha\beta}(y_0)y^{\alpha}y^{\beta}
-\frac{1}{6}R_{\alpha\beta;\gamma}
y^{\alpha}y^{\beta}y^{\gamma}+\left(\frac{1}{18}R_{\alpha\beta}
R_{\gamma\delta}-
\frac{1}{90}R_{\lambda\alpha\beta\kappa}
R^{\lambda\;\;\;\;\kappa}_{\;\;\gamma\delta}
+\frac{1}{20}R_{\alpha\beta;\gamma\delta}\right)
y^{\alpha}y^\beta y^\gamma y^\delta
+\Od(\partial^5)
\label{detmet}
\end{eqnarray}
For the square root of the metric determinant and its 
inverse we have:
\begin{eqnarray}
g^{1/2}(y)&=&1+\frac{1}{6}R_{\alpha\beta}(y_0)y^{\alpha}
y^{\beta}-\frac{1}{12}
R_{\alpha\beta;\gamma}
y^{\alpha}y^{\beta}y^{\gamma}
+\left(\frac{1}{72}R_{\alpha\beta}R_{\gamma\delta}-
\frac{1}{180}R_{\lambda\alpha\beta\kappa}
R^{\lambda\;\;\;\;\kappa}_{\;\;\gamma\delta}
+\frac{1}{40}R_{\alpha\beta;\gamma\delta}\right)
y^{\alpha}y^\beta y^\gamma y^\delta\nonumber \\
&+&\Od(\partial^5), 
\label{detmet12}\\
g^{-1/2}(y)&=&1-\frac{1}{6}R_{\alpha\beta}(y_0)y^{\alpha}
y^{\beta}+\frac{1}{12}
R_{\alpha\beta;\gamma}
y^{\alpha}y^{\beta}y^{\gamma}+\left(\frac{1}{72}R_{\alpha\beta}
R_{\gamma\delta}+
\frac{1}{180}R_{\lambda\alpha\beta\kappa}
R^{\lambda\;\;\;\;\kappa}_{\;\;\gamma\delta}
-\frac{1}{40}R_{\alpha\beta;\gamma\delta}\right)
y^{\alpha}y^\beta y^\gamma y^\delta\nonumber\\
&+&\Od(\partial^5)
\label{detmet-12}
\end{eqnarray}
\end{document}